\documentclass[12pt]{article}

\usepackage{amsmath,graphicx}
\usepackage[table]{xcolor}
\usepackage{fullpage}
\usepackage{setspace}
\usepackage{amsfonts}
\usepackage{lineno}
\usepackage{pdflscape}
\usepackage{hyperref}
\usepackage{authblk}
\usepackage{microtype}

\usepackage[
  backend=biber,
  bibstyle=vancouver,
  citestyle=numeric-comp,
  sorting=none,
  sortcites=true,
  doi=true,
  url=true,
  natbib=true
]{biblatex}
\title{\sc{Parameter Estimation for Differential Equation Models Using Generalized Profiling: A Computational Tutorial}}

\author[1,2]{Matthew J. Simpson}
\author[1,2]{James S. Bennett}
\author[3]{Ruth E. Baker}
\affil[1]{School of Mathematical Sciences, Queensland University of Technology (QUT), Brisbane, Australia.}
\affil[2]{ARC Centre of Excellence for the Mathematical Analysis of Cellular Systems, QUT, Brisbane,  Australia.}
\affil[3]{Mathematical Institute, University of Oxford, Oxford, OX2 6GG, United Kingdom.}

\date{}

\begin{document}

%\linenumbers

\maketitle

\begin{abstract}
Parameter estimation connects mathematical models to real-world data and decision making across many scientific and industrial applications. Standard approaches such as maximum likelihood estimation and Markov chain Monte Carlo estimate parameters by repeatedly solving the model, which often requires numerical solutions of differential equation models. In contrast, generalized profiling (also called parameter cascading) focuses directly on the governing differential equation(s), linking the model and data through a penalized likelihood that explicitly measures both the data fit and model fit. Despite several advantages, generalized profiling is not often used in practice. This tutorial-style article outlines a set of self-directed computational exercises that facilitate skills development in applying generalized profiling to a range of ordinary differential equation models. All calculations can be repeated using reproducible open-source Jupyter notebooks that are available on \href{https://github.com/ProfMJSimpson/PenalizedLikelihood}{GitHub}.
\end{abstract}

\newpage 
\section{Introduction} \label{sec:intro}

Parameter estimation is a critical step in the practical application of mathematical models, supporting fundamental scientific discovery and practical decision making across a  range of scientific and industrial applications~\cite{Simpson2026,Smith2014,Tarantola2005}.  Whether we are interested in forecasting the value of a share portfolio~\cite{Campbell1997}, predicting the motion of groundwater~\cite{Yeh1986}, or inferring neuronal connectivity from brain images~\cite{Friston2003}, a fundamental challenge in applied mathematical modeling is to determine values of model parameters so that the model solution best matches noisy, incomplete data and measurements.

Traditional approaches for parameter estimation, including likelihood-based~\cite{Hines2014,Pawitan2001,Wasserman2004} and likelihood-free~\cite{Gelman2013,Wilkinson2013ABC,Simpson2025Inference} methods, are very well established. Some scientific disciplines have developed in a way that focuses primarily on frequentist approaches by using optimization-based maximum likelihood estimation (MLE) to give point estimates of parameters.  In contrast, other disciplines have developed so that the predominant focus is on sampling-based Bayesian approaches, such as likelihood-based Markov chain Monte Carlo (MCMC) or likelihood-free Approximate Bayesian computation (ABC), to target the full posterior parameter distribution. Regardless of which standard approach is taken, their application to ODE-based mathematical models requires having access to the solution of the ODE associated with the mathematical model.  The solution could be an exact analytical solution, or it could be an approximate numerical solution. This requirement of standard approaches introduces certain considerations that motivate the following questions:

\begin{enumerate}\itemsep = 1mm
\item Is it possible to derive an exact solution of the ODE, or are we limited to working with approximate numerical solutions? 
\item When working with numerical solutions, what is the computational overhead involved in solving the ODE numerically?
\item When working with numerical solutions, what is the impact of numerical truncation error on parameter estimation accuracy?
\item When we evaluate the exact or numerical solution, are there special cases to consider when evaluating these solutions across the parameter space (e.g.~see Section~\ref{sec: CRN})? 
\item How do we estimate the initial condition(s) required to solve the ODE(s)? 
\end{enumerate}

Instead of addressing these questions associated with standard parameter estimation methods, alternative strategies that avoid these questions can be adopted. Generalized profiling~\citep{Ramsay2007,Ramsay_Hooker_2017}, also known as parameter cascading or smoothing~\citep{Cao2007,Cao2008}, is fundamentally different because it involves proposing ``trial functions'' that approximately satisfy the governing ODEs rather than working with their solution. While generalized profiling has been adopted within a relatively small number of application areas, it has not enjoyed the same widespread deployment as the more standard approaches described above.  In brief, generalized profiling involves approximating the ODE solution using a smooth function that is chosen to approximately fit the data while also approximately enforcing the ODE. Model parameters are inferred by adjusting this function and the model parameters together, to strike a balance  between the fit of the function to the data and enforcement of the ODE. 

The aim of this article is to provide a series of practical tutorial materials and computational exercises that focus on implementing a straightforward form of generalized profiling.  Computational resources are provided in an open source notebook format, with software written in Julia~\citep{Bezanson2017}.  This tutorial material is especially timely, given the recent surge of interest in using Physics-Informed Neural Networks (PINNs)~\citep{Raissi2019,Tartakovsky2020} and Biologically-Informed Neural Networks (BINNs)~\citep{Lagergren2020BINNs} for parameter estimation.  A key component of PINNs/BINNs-based approaches is to evaluate \textit{data loss} and \textit{physics loss} terms that are directly analogous with the \textit{data matching} and \textit{model matching} terms in generalized profiling. The recent surge in using PINNS/BINNs-based approaches appears to have largely overlooked the deeper connection with ideas previously established in the generalized profiling literature.

This article is structured in the following way. To set the scene, we briefly explore data interpolation and data approximation using B-splines. This material explains how B-splines can be used to approximate functions that describe noisy data, as well as computing their derivatives, and they will be the trial functions that we employ. With these interpolation tools, we then introduce measures of data matching and model matching that provide the foundation to combine tools from computational linear algebra and numerical optimization to estimate parameters of the ODE-based models. These parameter estimates are iteratively updated to strike a balance between having a spline representation of the ODE solution that approximately matches the data, while also approximately enforcing the relevant ODE. We work with a number of illustrative examples, using both synthetic and real-world data, covering models based on both scalar ODEs and systems of ODEs.

\section{Background: Function approximation with B-Splines} \label{sec:Spline}

\begin{figure}[htp]
  \centering
\includegraphics[width=1.0\textwidth]{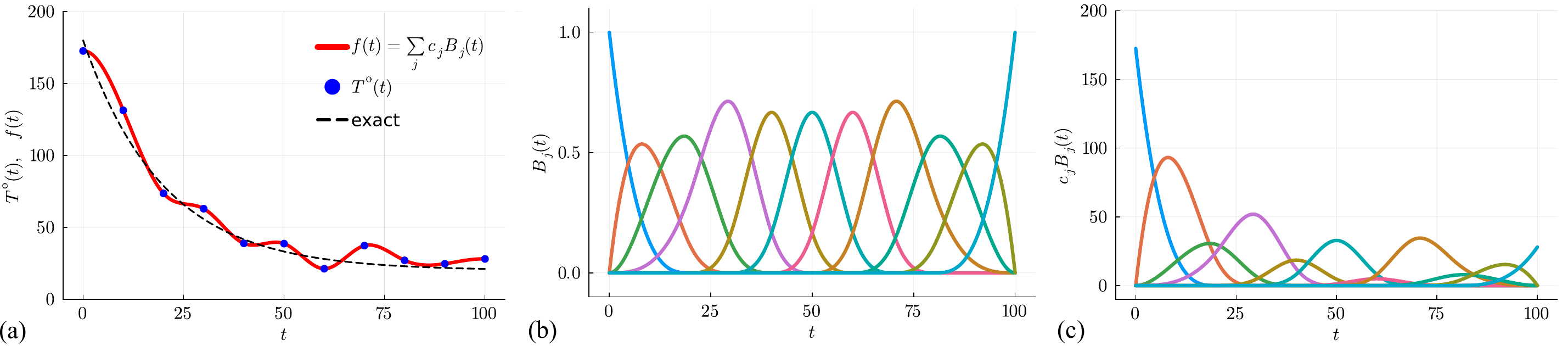}
  \caption{(a) Dashed curve shows $g(t) = 20 + 160 \, \textrm{exp}(-t/20)$ on  $t \in [0, 100]$. Blue dots show $J=11$ equally-spaced noisy samples at $t=0,10,20,\ldots,100$.  The $j$th data point is given by  $T^{\textrm{o}}(t_j) = g(t_j) + \mathcal N(0,\sigma^2)$ for $j=1,2,3,\ldots,J$, with $\sigma=8$ for this data realization. The red curve is a cubic B-spline interpolation.  (b)  Unscaled cubic basis functions, $B_j(t)$ for $j=1,2,3,\ldots,J$. (c) Scaled cubic basis functions, $c_j B_j(t)$ for $j=1,2,3,\ldots,J$. The colors used to plot $B_j(t)$ in (b) are the same as the colors used to plot the corresponding $c_j B_j(t)$ in (c). Summing the scaled cubic basis functions in (c) gives the B-spline interpolation in (a), $f(t) = \sum_{j=1}^{J} c_j B_j(t)$ (red curve). \label{fig:F1}}
\end{figure}

A key step in generalized profiling is to approximate a set of noisy data, such as the data in Figure~\ref{fig:F1}(a).  To motivate this, consider an exponentially decreasing function of time,  $g(t) = 20 + 160 \exp(-t/20)$ for  $t \in [0,100]$.  To generate noisy data, we evaluate $g(t)$ at 11 equally-spaced points in time, and introduce additive Gaussian noise so that the $j$th data point takes the form $T^{\textrm{o}}(t_j) = g(t_j) + \mathcal N(0,\sigma^2)$ for $j=1,2,3,\ldots, J$, where $J=11$\footnotemark[4]\footnotetext[4]{We use the symbol $\sigma^2$  to denote variance in the usual way. Our software refers to $\sigma$ only to be consistent with the syntax requirements of the Distributions.jl package for Julia.}.   The $j$th data point is denoted $T^{\textrm{o}}(t_j)$, with the superscript `o' indicating this is an \textit{observed} quantity, and we use the symbol $T$ because later in Section~\ref{sec: Newton} we will re-visit this data to calibrate a mathematical model describing temperature.  Roughly speaking, the data in Figure~\ref{fig:F1}(a) shows an overall decreasing trend for $t \in [0, 100]$.  The impact of the additive Gaussian noise is clear since subsequent data points fluctuate randomly about the underlying function $g(t)$.

We approximate discrete data using \textit{basis splines}, also known as B-splines.  All calculations are performed using the open access Julia package \verb |BSplineKit.jl|~\citep{Polanco2025BSplineKit}, and we now outline key features about B-splines in a brief, implementation-focused summary.  First described in 1946 by Schoenberg~\cite{Schoenberg1946}, additional information about B-splines can be found in de Boor's monograph~\citep{deboor2001} and the review by Botella and Shariff~\citep{Botella2003}.  In this work we use B-splines to both interpolate noisy data and to approximate a function that best-describes the solution of an ODE model that is thought to describe that noisy data.  Therefore, we treat B-splines as a tool for data interpolation and data approximation, making a clear distinction between these two roles where relevant.  

A B-spline of order $p+1$ is a piecewise polynomial, where each ``piece'' consists of a polynomial of degree $p \in \mathbb{Z}_{\ge 0} = \{0,1,2,\dots\}$. For example, a B-spline of order four is composed of piecewise of cubic polynomials and $p=3$. The polynomial pieces  meet at special points called \textit{knots}~\citep{deboor2001,Botella2003}. The continuity and derivative requirements of the function being interpolated determine how often each knot is repeated, which is  called the knot \textit{multiplicity}~\citep{deboor2001,Botella2003}.  We will discuss why knots are repeated and the impact of their repetition in Section~\ref{sec: Newton}.  An important consideration is the role of boundary (end-point) conditions, with the most straightforward option of not imposing any constraints on the B-spline at the end points of the interval.  This is sometimes called \textit{unconstrained} boundary conditions and is the default option in \verb |BSplineKit.jl|~\citep{Polanco2025BSplineKit}.  Another option is to implement \textit{natural} boundary conditions which requires the second derivative of the B-spline to vanish at the end points of the interval~\citep{deboor2001,Botella2003}.  All results presented here involve standard unconstrained boundary conditions, however, the software we provide on GitHub can be easily modified to repeat the calculations using natural boundary conditions.  We find that our results are relatively insensitive to this change and we encourage readers to explore this by making minor modifications to the software, as described in the Jupyter notebooks.

We now outline the concepts, definitions and notation required to develop B-spline tools.  B-spline notation can be developed using degree-based or order-based frameworks.  Here we work with degree-based notation because this is consistent with the conventions adopted in \verb |BSplineKit.jl|~\citep{Polanco2025BSplineKit}.  A B-spline of order $p+1$ can be specified using a non-decreasing knot sequence
\begin{equation}
t_0 \le t_1 \le \cdots \le t_m,
\end{equation}
where some knots may be repeated.  The associated B-spline basis functions are denoted $B_{j,p}(t)$, where $j$ indexes the basis function. In generalized profiling, the first step is to interpolate noisy data (e.g., Figure~\ref{fig:F1}(a)) using $J$ basis functions $B_{j,p}(t)$ for $j=1,\ldots,J$~\citep{deboor2001,Botella2003}. We note that B-splines have local support: $\operatorname{supp}(B_{j,p}) \subseteq [t_j,\, t_{j+p+1})$, so $B_{j,p}(t)=0$ for $t\notin [t_j,\, t_{j+p+1})$.

Basis functions are defined recursively beginning with piecewise constant ($p=0$) functions
\begin{equation}
B_{j,0}(t) =
\begin{cases}
1, &  t \in [t_j,t_{j+1}),\\
0, & t \notin [t_j,t_{j+1}).
\end{cases}
\end{equation}
For $p\ge 1$  the Cox--de Boor formula~\cite{Cox1972NumericalEvaluationBSplines,deBoor1972CalculatingBSplines} gives higher degree basis functions as follows,
\begin{equation}
B_{j,p}(t)
= \left[\dfrac{t-t_j}{t_{j+p}-t_j}\right]\,B_{j,p-1}(t)
+
\left[\dfrac{t_{j+p+1}-t}{t_{j+p+1}-t_{j+1}}\right]\,B_{j+1,p-1}(t).
\end{equation}
For the recursion, we adopt the convention that any term with a zero denominator is set to zero, which is essential for handling repeated knots. We focus on cubic B-splines with $p=3$ and, going forward, we write $B_j(t)$ in place of $B_{j,3}(t)$. Each cubic basis function depends only on the five knots $(t_j,\ldots,t_{j+4})$ and has support on $[t_j,t_{j+4})$, so $B_j(t)=0$ for $t\notin[t_j,t_{j+4})$.

With $J=11$ data points in Figure~\ref{fig:F1}(a), we use \verb|BSplineKit.jl| to obtain basis functions $B_{j}(t)$ for $j=1,2,3,\ldots,J$.  Since our data are collected at $t=0,10,20,\ldots, 100$, and we choose to work with cubic splines, \verb|BSplineKit.jl| places knots at 
\begin{equation*}
\mathbf{t} = (0,0,0,0,20,30,40,50,60,70,80,100,100,100,100)^\top,
\end{equation*}
noting that knots at $t=0$ and $t=100$ have multiplicity four because with unconstrained boundary conditions boundary knots have multiplicity equal to the spline order.  In general, the number of knots is given by the sum of the number of splines and the order, in this case $11 + 4 = 15$. Since we have four knots at $t=0$ and four knots at $t=100$, we are left with  $15-2\times4 = 7$ interior knots that \verb|BSplineKit.jl| places at $(20,30,40,50,60,70,80)^\top$. For cubic splines with unconstrained boundary conditions it is typical to omit knots at the second and second-to-last data points~\citep{deboor2001,Botella2003}, which are $t=10$ and $t=90$ in this case. While \verb|BSplineKit.jl| adopts this convention, it is possible to omit different knots.

The $J=11$ basis functions in Figure~\ref{fig:F1}(b) are cubic polynomials with support $[t_j,t_{j+4})$. For example, $B_1(t)$ is supported on $[0,20)$ with $B_1(0)=1$, while $B_2(t)$ is supported on $[0,30)$ and includes a turning point. Plotted together, we see the basis functions are symmetric about $t=50$---this is because the data are equally spaced. The central basis functions ($B_j(t)$ for $j=5,6,7$) are identical up to translation, and the boundary basis functions ($B_j(t)$ for $j=1,\ldots,4$ and $j=8,\ldots,11$) differ from each other, and mirror each other about the midpoint of the interval, $t=50$.

A key feature of B-splines is that any spline of order $p+1$ defined on a given knot sequence can be written as a linear combination of the B-spline basis functions of order $p+1$ associated with that knot sequence.  This means that we can interpolate the noisy data using a spline representation as follows:
\begin{equation}\label{eq:Spline}
f(t)
=
\sum_{j=1}^{J} c_j B_j(t),
\end{equation}
where $f(t)$ is the spline representation of the noisy data. The coefficients $c_j$ for $j=1,\ldots,J$ are chosen so that $f(t)$ interpolates the data, yielding a linear system for the coefficients that is solved in \verb|BSplineKit.jl|~\citep{Polanco2025BSplineKit}. Figure~\ref{fig:F1}(c) shows the scaled B-splines $c_jB_j(t)$: their shapes match those in Figure~\ref{fig:F1}(b), but are vertically scaled according to $c_j$. For example, $c_1=172.5296$ scales the left-most basis function by a factor of $172.5296$.

Summing the scaled basis functions in Figure~\ref{fig:F1}(c) gives the spline interpolation $f(t) = \sum_{j=1}^{J} c_j B_{j}(t)$ that is shown in Figure~\ref{fig:F1}(a).  This spline is superimposed on the noisy data, and on the original function, $g(t)$.  The B-spline provides a smooth interpolation that passes through all data points, which is the first step in generalized profiling.  We will refer to this initial spline as \textit{over-fitting} the data because it fits the noise rather than the underlying signal.  As we will subsequently explain, further iterations of generalized profiling will use the governing ODE to adapt the spline so that it approximates the underlying signal rather than the noise. 

\begin{figure}[htp]
  \centering
\includegraphics[width=1.0\textwidth]{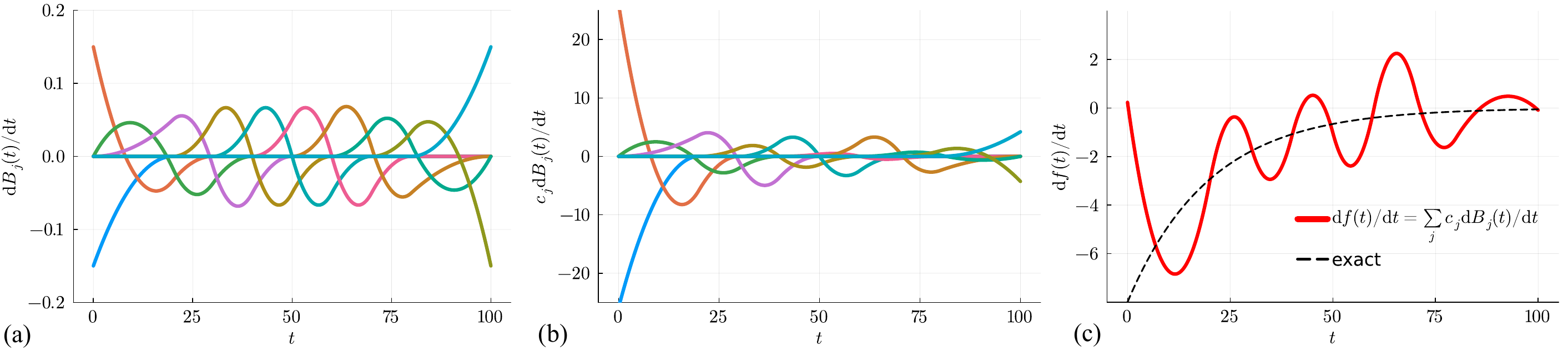}
  \caption{(a)  Plots of $\textrm{d} B_j(t) / \textrm{d} t$ for $j=1,2,3,\ldots, J$ based on the unscaled cubic basis functions in Figure~\ref{fig:F1}(b).  The color of each $\textrm{d} B_j(t) / \textrm{d} t$ corresponds to the colors in Figure~\ref{fig:F1}(b) for each cubic basis function, $B_j(t)$ for $j=1,2,3,\ldots, J$.   (b)  Scaled quadratic polynomials $c_j \textrm{d} B_j(t) / \textrm{d} t$ for $j=1,2,3,\ldots,J$. Each scaled quadratic polynomial is plotted using the same color as the corresponding unscaled quadratic polynomial in (a).  Summing the scaled piecewise quadratic polynomials in (b) gives the red curve in (c),  $\textrm{d}f(t)/ \textrm{d}t   = \sum_{j=1}^{J} c_j \, \textrm{d}B_j(t) /\textrm{d}t$.  This approximation is superimposed on the exact result  $\textrm{d}g(t)/\textrm{d}t = -8\exp(-t/20)$. For this data set we have $J=11$. \label{fig:F2}}
\end{figure}

Since our prime focus is to use the B-spline polynomials to estimate parameter values in an ODE model, it is relevant to evaluate $\textrm{d}f(t) / \textrm{d}t$. Figure~\ref{fig:F2}(a) shows $\textrm{d} B_j(t) / \textrm{d}t$ for $j=1,2,3,\ldots, J$, which are a family of piecewise quadratic polynomials obtained by differentiating the basis functions in Figure~\ref{fig:F1}(a). Scaling these piecewise quadratic polynomials using the same coefficients, $c_j$ for $j=1,2,3,\ldots, J$,  gives the scaled piecewise quadratic polynomials in Figure~\ref{fig:F2}(b). Summing the scaled piecewise quadratic polynomials gives 
\begin{equation}
\dfrac{\textrm{d} f(t)}{\textrm{d} t} = \sum_{j=1}^{J} c_j \, \dfrac{\textrm{d}B_j(t)}{\textrm{d}t},
\end{equation}
and  Figure~\ref{fig:F2}(c) compares $\textrm{d} f(t) / \textrm{d} t$ and  $\textrm{d}g(t) / \textrm{d}t$.  Here we see that the spline approximation smoothly fluctuates about the exact result, but overall it provides a reasonable approximation of the first derivative across the interval of interest. 

Before introducing a parameter estimation example, we define collocation matrices. For basis functions $B_j(t)$ ($j=1,\ldots,J$), let $\mathbf{A}$ be the $I\times J$ matrix with entries $A_{i,j}=B_j(t_i)$ ($i=1,\ldots,I$). We also define $\mathbf{A}'$ with entries $A'_{i,j}=\mathrm{d}B_j(t_i)/\mathrm{d}t$, also of dimensions $I\times J$. These matrices are banded and sparse, so they can be formed and stored efficiently~\citep{Botella2003}. Students may have encountered collocation matrices in advanced numerical methods courses that cover finite element analysis~\citep{Segerlind1976}.

\section{Parameter estimation for ODEs}

We now explore three case studies (Sections~\ref{sec: Newton}--\ref{sec: CRN}) using well-defined synthetic data to build skills in generalized profiling before deploying these skills on practical data set from  population biology (Section~\ref{sec:Coral}).

\subsection{Newton's law of cooling}\label{sec: Newton}

The noisy data in Figure~\ref{fig:F1}(a) was carefully chosen so that we can re-visit this data to calibrate a mathematical model.  We will interpret this data using a very simple ODE model that is often encountered in undergraduate courses on mathematical modeling.  The mathematical model describes the cooling (or heating) of an object at uniform temperature $T(t)$, where the temperature can vary with time.  The object, initially at some temperature $T(0)$, is placed in an environment of constant ambient temperature, $T_{\textrm{a}}$.  Heat conduction leads to $T(t)$ decreasing with time if $T(0) > T_{\textrm{a}}$, or $T(t)$ increasing if $T(0) < T_{\textrm{a}}$.  This heat transfer is often modeled using Newton's law of cooling~\citep{Simpson2026} 
\begin{equation}
\label{eq:NewtonODE}
\dfrac{\textrm{d}T(t)}{\textrm{d}t}=- \alpha \left(T(t) - T_{\rm{a}}\right), 
\end{equation}
where $\alpha > 0$ is a constant heat transfer coefficient.  Before applying generalized profiling to calibrate Newton's law of cooling, it is useful to recall key steps in a more traditional approach.  The solution of Equation \eqref{eq:NewtonODE} is 
\begin{equation}
T(t) = \left(T(0)-T_{\textrm{a}}\right)\textrm{e}^{-\alpha t} + T_{\textrm{a}},
\end{equation}
confirming that we must have estimates of $T(0)$, $\alpha$ and  $T_{\textrm{a}}$ to evaluate the solution $T(t)$.  A notable feature of the data in Figure~\ref{fig:F1}(a) is that it is noisy, and the magnitude of the noise is controlled by the parameter $\sigma$, which can also be treated as an unknown quantity to be estimated~\cite{Simpson2022}.  Unknown parameters are typically collected in a vector $\boldsymbol{\theta}_{\textrm{full}} = (T(0),\alpha,T_{\textrm{a}},\sigma)^\top$, here we use the subscript `full' to indicate that this is the \textit{full} vector of parameters to distinguish this parameter grouping from the reduced set of parameters that we will work with under generalized profiling below.  These four parameters could be estimated in a more traditional framework using MLE~\citep{Simpson2026} or MCMC to generate a posterior distribution~\citep{Hines2014} from which point estimates of the parameters for which the solution of the ODE gives the best match to the data can be obtained, $\hat{\boldsymbol{\theta}}_{\textrm{full}} = (\hat{T}(0),\hat{\alpha},\hat{T_{\textrm{a}}},\hat{\sigma})^\top$.  

Before proceeding to estimate $\boldsymbol{\theta}$, we note that the data realization in Figure~\ref{fig:F1}(a) is generated using the \textit{true} parameters $(T(0),\alpha,T_{\textrm{a}},\sigma)^\top = (180,0.05,20,8)^\top$, and we now estimate these parameters using generalized profiling. At this early stage we see a clear advantage of generalized profiling: we do not need to solve the ODE model which means that we do not need to estimate $T(0)$.  In contrast, standard MLE or MCMC-based approaches require estimates of $T(0)$ to evaluate the ODE solution.  Instead, here we estimate $\boldsymbol{\theta} = (\alpha,T_{\textrm{a}},\sigma)^\top$ using generalized profiling, and then estimate $T(0)$ as a simple post-processing step.  The computational saving of not estimating $T(0)$ is modest for a scalar ODE, but potentially substantial for ODE systems (see Section~\ref{sec: CRN}).

\begin{figure}[htp]
  \centering
\includegraphics[width=1.0\textwidth]{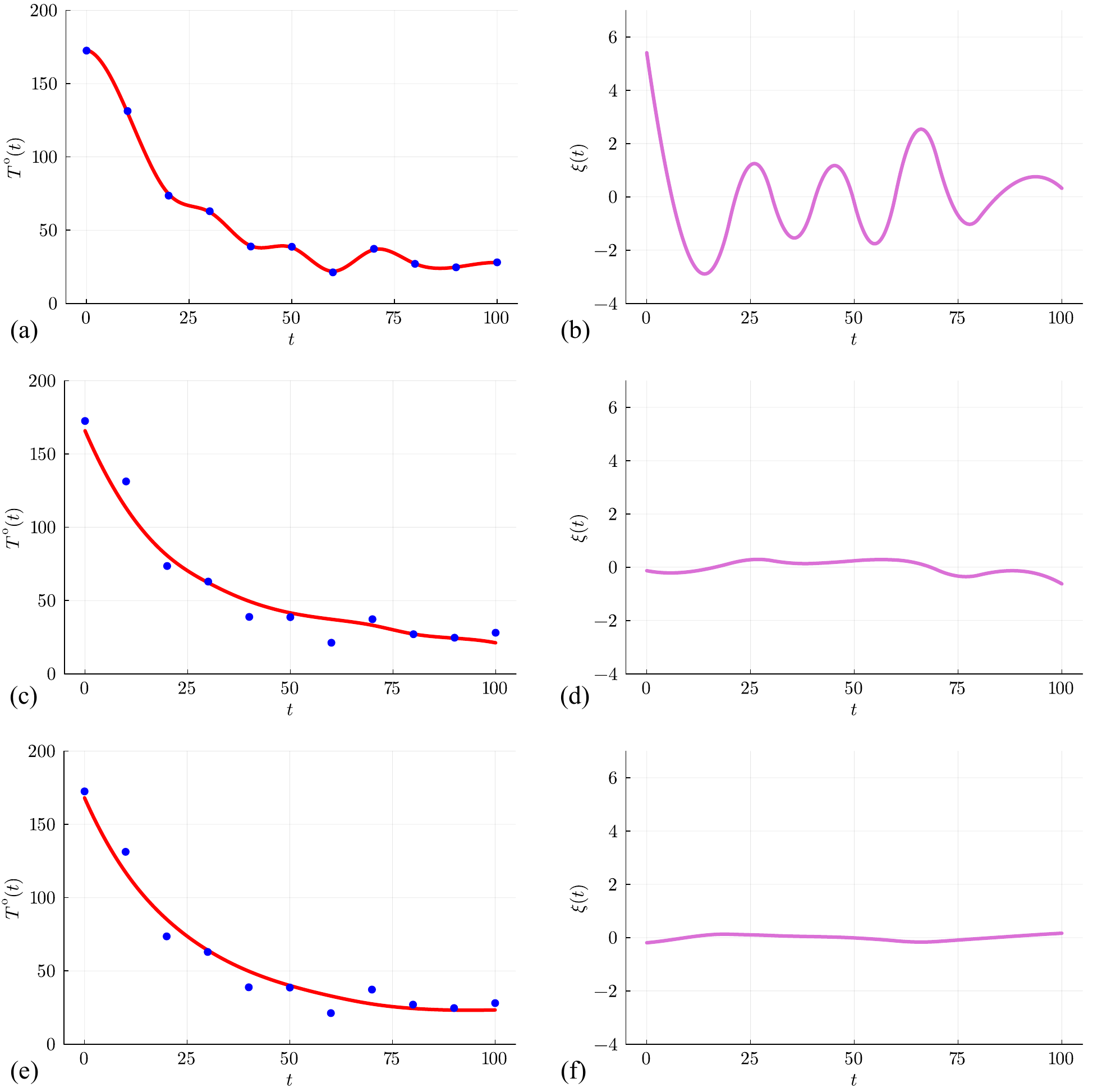}
  \caption{Newton's law of cooling case study with synthetic data. (a) Data (blue dots) and over-fitted spline (red curve) from Figure~\ref{fig:F1}(a). (b) Plot of $\xi(t;\boldsymbol{\theta}^{(1)})$ (purple curve) with $\boldsymbol{\theta}^{(1)} = \left(\alpha, T_{\textrm{a}},\sigma \right)^\top = \left(0.037, 17.229, \sigma  \right)^\top$.  Results in (c)--(d) illustrate the outcome after two iterations, with (c) showing the updated spline (red curve) superimposed on the data (blue dots) and (d) showing $\xi(t;\boldsymbol{\theta}^{(2)})$ (purple curve) with $\boldsymbol{\theta}^{(2)} = (\alpha, T_{\textrm{a}},\sigma)^\top= (0.044, 23.1100, 8.874)^\top$. Results in (e)--(f) illustrate the outcome after ten iterations, with (e) showing the updated spline (red curve) superimposed on the data (blue dots) and (f) showing $\xi(t;\boldsymbol{\theta}^{(10)})$ (purple curve) with $\boldsymbol{\theta}^{(10)} = (\alpha, T_{\textrm{a}},\sigma)^\top = (0.042, 20.833, 7.737)^\top$. Calculations are performed using $J=11$ and $K=1001$.\label{fig:F3}}
\end{figure}

The first step in generalized profiling is to use B-splines (Section~\ref{sec:Spline}) to interpolate the data, as illustrated in Figure~\ref{fig:F3}(a) where we plot the data and a B-spline that interpolates it.  This spline ``overfits'' the data in the sense that we do not expect the true solution of the ODE model to go through all the data points. This spline gives a continuous approximation of both $T(t)$ and $\textrm{d}T(t) / \textrm{d}t$, which we write as $f(t)$ and  $\textrm{d}f(t) / \textrm{d}t$, respectively.  With these quantities we compute a new continuous function
\begin{equation}
\xi(t; \boldsymbol{\theta}) = \dfrac{\textrm{d}f(t)}{\textrm{d}t}  -\Bigg[ -\alpha \left(f(t) - T_{\rm{a}}\right) \Bigg],
\end{equation}
which we  interpret as a measure of \textit{discrepancy}, noting that if $f(t)$ satisfies Equation~\eqref{eq:NewtonODE} then $\xi(t; \boldsymbol{\theta}) \equiv 0$.   At this point, our aim is to determine the values of $T_{\textrm{a}}$ and $\alpha$ that lead to $\xi(t;\boldsymbol{\theta})$ being as close to zero as possible.  To implement this we consider a fine uniform discretisation of the independent variable, $t_k = (k-1)/10$, for $k=1,2,3,\ldots,K$, where $K=1001$, and compute the quantity
\begin{equation}
\ell_{\textrm{m}}\left(\boldsymbol{\theta}; T^{\textrm{o}}(t) \right) = -\dfrac{1}{K}\sum_{k=1}^{K}\left[\xi(t_k;\boldsymbol{\theta})\right]^2, \label{eq:modelerror}
\end{equation}
where the subscript `m' indicates that $\ell_{\textrm{m}}$ measures of how well $f(t)$ satisfies the ODE model.  

Using numerical optimization, and given the spline function $f(t)$, we estimate the $\boldsymbol{\theta}$ that maximizes $\ell_{\textrm{m}}$.  All numerical optimization calculations in this work are computed with the Nelder--Mead algorithm using simple bound constraints within the NLopt library of optimization routines~\citep{Johnson2024}. The Nelder--Mead algorithm is a standard choice for numerical optimization that is extensively tested and well understood~\citep{Audet,Nocedal}.  In this case we obtain $\boldsymbol{\theta}^{(1)} = \left(\alpha, T_{\textrm{a}},\sigma \right)^\top = \left(0.037, 17.229, \sigma  \right)^\top$, noting that $\sigma$ plays no role here since $\ell_{\textrm{m}}$ is independent of $\sigma$, and we will explain how to estimate $\sigma$ in subsequent steps. With these values of $\alpha$ and $T_{\textrm{a}}$ we plot $\xi(t; \boldsymbol{\theta}^{(1)})$ in Figure~\ref{fig:F3}(b) which shows that these initial estimates, generated from the over-fitted spline, are far from satisfying $\xi(t; \boldsymbol{\theta})=0$, indicating that the spline in Figure~\ref{fig:F3}(a) does not satisfy Equation~\eqref{eq:NewtonODE}.

The second step involves updating the spline coefficients to balance the competing requirements that the spline should pass sufficiently close to the data points while also approximately satisfying the ODE.  We achieve this by forming a linear least--squares problem that we solve using Julia's backslash operator~\citep{Bezanson2017}. In particular, we form an over-determined \textit{stacked} linear system
\begin{equation} \label{eq:LeastSquares}
\begin{bmatrix}
w_\textrm{d} \mathbf{A} \\[0.5em]
w_\textrm{m} \mathbf{A}'
\end{bmatrix} \mathbf{c}
=
\begin{bmatrix}
w_\textrm{d} \mathbf{b} \\[0.5em]
w_\textrm{m} \mathbf{b}'
\end{bmatrix},
\end{equation}
where $\mathbf{A}$ is a square collocation matrix of size $J \times J$ with entries $B_j(t_j)$, $\mathbf{A}'$ is a tall collocation matrix of size $K \times J$ with entries $\textrm{d}B_j(t_k)/\textrm{d}t$, $\mathbf{b}$ is a column vector of length $J$ containing the data, $T^{\textrm{o}}(t_{j})$, $\mathbf{b}'$ is a column vector of length $K$ with entries corresponding to the right-hand side of the ODE written in terms of the spline, $-\alpha\left(f(t_k) - T_{\textrm{a}} \right)$ for $k=1,2,3,\ldots, K$, and $\mathbf{c}$ is a column vector of length $J$ containing the spline coefficients.  Since $J=11$ and $K=1001$, the stacked matrix has size $1012 \times 11$, the stacked vector has length $1012$, and the coefficient vector $\mathbf{c}$ has length $11$.

The first $J$ rows in Equation~\eqref{eq:LeastSquares} are associated with \textit{data} matching and the remaining $K$ rows are associated with \textit{model} matching. Note also that we introduce a vector of positive weights, $\mathbf{w} = (w_\textrm{d},w_\textrm{m})^\top$, for which the subscript `d' indicates that $w_\textrm{d}$ is a \textit{data} weight, and the subscript `m' indicates that $w_\textrm{m}$ is a \textit{model} weight. These two weights balance the emphasis between the spline solution matching the data and satisfying the ODE. In the first iteration we set $w_\textrm{d} = 1$ and $w_\textrm{m} = | 1/\ell_{\textrm{m}}|$. The solution of Equation~\eqref{eq:LeastSquares} updates $c_j$ for $j=1,2,3,\ldots,J$, giving an updated spline shown in Figure~\ref{fig:F3}(c). The updated spline is a ``smoothing'' spline that fits the data in a least--squares sense without being constrained to pass through each data point owing to the impact of the model matching terms in the final $K$ rows of the stacked system.  From this point forward the spline takes on a role as means of approximating rather than interpolating the data.

To update $\boldsymbol{\theta}$ we use Equation~\eqref{eq:modelerror} to estimate $\ell_{\textrm{m}}(\boldsymbol{\theta};T^{\textrm{o}}(t))$, as well as calculating a measure of distance between the updated spline and the data.  We work with the standard assumption that the data are normally distributed about the spline with constant variance, $\sigma^2$.  Invoking a regular independence assumption, this framework leads to the following expression for the log-likelihood~\citep{Simpson2026}:
\begin{equation}
\ell_{\textrm{d}}\left(\boldsymbol{\theta}; T^{\textrm{o}}(t) \right) = \sum_{j=1}^{J} \log \left[\phi\left(T^{\textrm{o}}(t_j); f(t_j), \sigma^2  \right) \right],
\end{equation}
where $\phi(x; \mu, \sigma^2)$ is the density of the normal distribution with mean $\mu$ and variance $\sigma^2$.  The subscript `d' indicates that $\ell_{\textrm{d}}$ measures how well the spline matches the data.  Here we proceed using this choice of $\ell_{\textrm{d}}$, but we note that there are different choices that could be made, such as working with a least--squares definition or working with a different noise model (see Section \ref{sec: CRN}).  With these ingredients we form a penalized log-likelihood function which is a weighted sum of the standard loglikelihood function for the data $\ell_{\textrm{d}}$, and $\ell_{\textrm{m}}$ which measures how well $f(t)$ satisfies the ODE model. Together this gives
\begin{equation}\label{eq: Penalised Log Likelihood}
\ell(\boldsymbol{\theta}; T^{\textrm{o}}(t)) = w_\textrm{d} \ell_{\textrm{d}}(\boldsymbol{\theta}; T^{\textrm{o}}(t))+w_\textrm{m} \ell_{\textrm{m}}(\boldsymbol{\theta}; T^{\textrm{o}}(t)).
\end{equation}
The two weights are determined iteratively as follows.  Given estimates of $\ell_{\textrm{d}} < 0$ and $\ell_{\textrm{m}} < 0$, we set $w_{\textrm{d}} = |1/\ell_{\textrm{d}}|$ and $w_\textrm{m} =| 1/\ell_{\textrm{m}}|$ in Equations~\eqref{eq:LeastSquares} and \eqref{eq: Penalised Log Likelihood} to balance the requirement that the spline ought to pass sufficiently close to the data while also sufficiently satisfying the ODE~\cite{Wu2023LikelihoodBasedMeasles}.  Splines that provide a better match to the data are associated with larger values of $w_\textrm{d}$ whereas splines that provide a better match to the model are associated with larger values of $w_\textrm{m}$.  Using numerical optimization we compute values of $\boldsymbol{\theta}$ that maximize $\ell$, giving $\boldsymbol{\theta}^{(2)} = (\alpha, T_{\textrm{a}},\sigma)^\top= (0.044, 23.110, 8.874)^\top$ with which we compute $\xi(t; \boldsymbol{\theta}^{(2)})$.  This updated discrepancy is plotted in Figure~\ref{fig:F3}(d), indicating that the updated spline enforces the ODE more accurately since $\xi(t,\boldsymbol{\theta}^{(2)})$ has moved closer to zero relative to $\xi(t,\boldsymbol{\theta}^{(1)})$.

Given  $\boldsymbol{\theta}^{(2)}$, we  iteratively repeat the three step process of sequentially updating:
\begin{enumerate}\itemsep=1mm
\item the spline coefficients, $c_j$ for $j=1,2,3,\ldots,J$, by solving Equation~\eqref{eq:LeastSquares},
\item the weights, $w_\textrm{d} = |1/\ell_{\textrm{d}}|$ and $w_\textrm{m} =| 1/\ell_{\textrm{m}}|$; and
\item the parameters $\boldsymbol{\theta}$, using numerical optimization to maximize Equation~\eqref{eq: Penalised Log Likelihood}.
\end{enumerate}
With each iteration we visually observe that the spline approaches a monotonically decreasing function while $\xi(t;\boldsymbol{\theta})$ approaches zero, indicating that the spline approximately satisfies Equation~\eqref{eq:NewtonODE}.  After 10 iterations we obtain the spline in Figure~\ref{fig:F3}(e) and the discrepancy in Figure~\ref{fig:F3}(f), giving $\boldsymbol{\theta}^{(10)} = (\alpha, T_{\textrm{a}},\sigma)^\top = (0.042, 20.833, 7.737)^\top$, which provides a reasonable approximation of the true values. Note that using our spline representation of $T(t)$ we can evaluate $f(0)$ to give an estimate of $T(0)$. In this case we have $f(0) = 167.891$, while the true value is $T(0) = 180$.

\subsection{Logistic growth}\label{sec: Logistic}

To reinforce the skills developed in Section~\ref{sec: Newton}, we apply the same method to a different scalar ODE.  The logistic growth  model~\citep{murray2002,Tsoularis2002} can be written as,
\begin{equation}
\dfrac{\textrm{d}C(t)}{\textrm{d}t}=\lambda C(t)\left[1-\dfrac{C(t)}{\kappa}\right], \label{eq:LogisticODE}
\end{equation}
where $C(t)\ge 0$ is population density, $\lambda>0$ is the growth rate, and $\kappa>0$ is the carrying capacity. For $C(0)/\kappa\ll 1$, growth is initially exponential, then slows as $C(t)\to\kappa^-$ due to competition (e.g., for space or nutrients), producing the familiar sigmoid growth curve~\citep{murray2002}. The logistic growth model is widely used across a wide range of problems that range from modeling microscopic cell populations over several hours~\citep{Maini2004} to modelling the growth of polyp populations on the surface of oyster shells over a month~\cite{Melica2014} right up to modeling continental-scale human population growth over millennia~\citep{Steele1998}. 

\begin{figure}[htp]
  \centering
\includegraphics[width=1.0\textwidth]{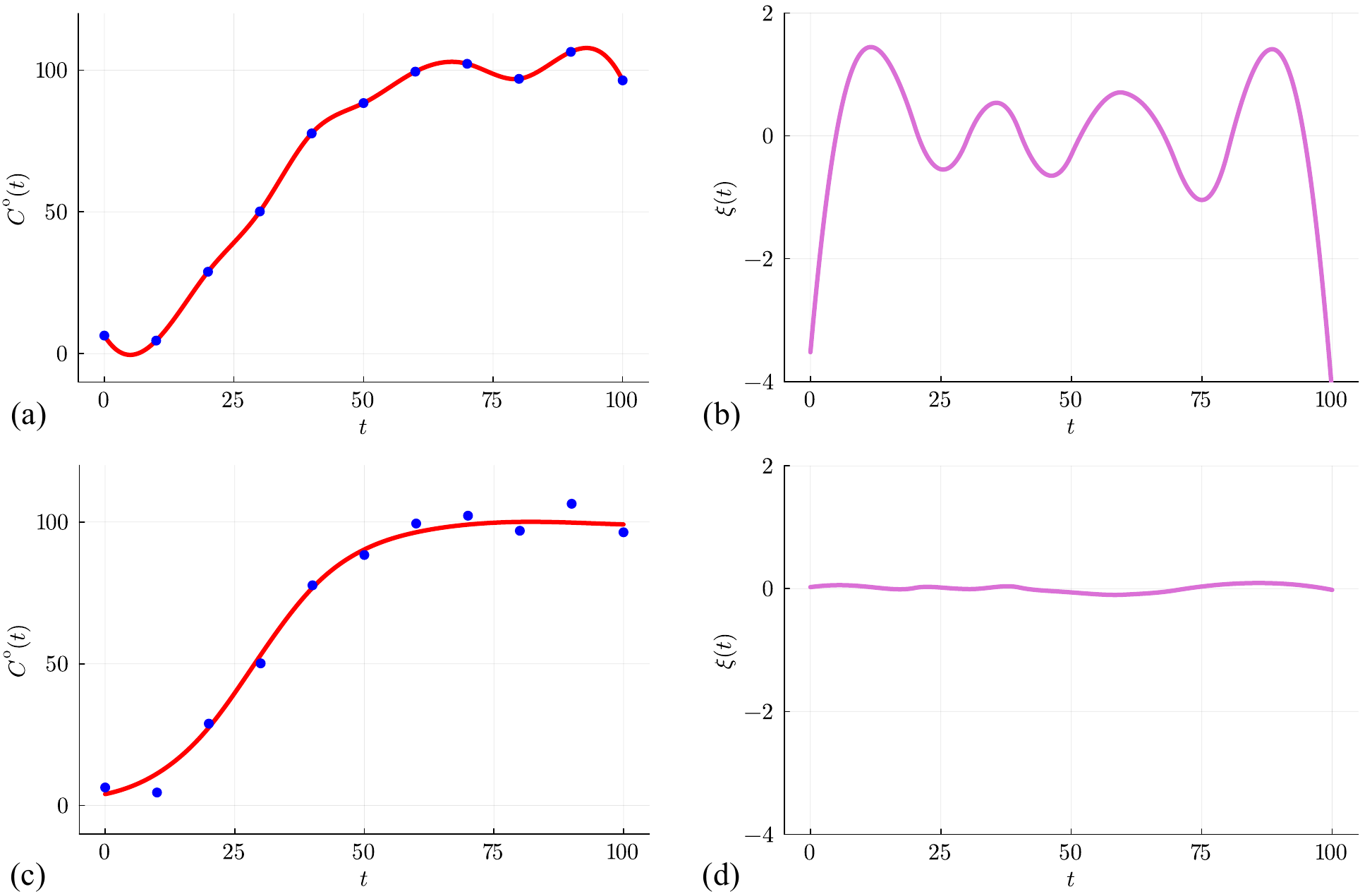}
  \caption{Logistic growth case study with synthetic data. (a) Data (blue dots) are collected at $t=0, 10, 20,\ldots, 100$ and interpolated using an over-fitted spline (red curve). (b) Plot of $\xi(t;\boldsymbol{\theta}^{(1)})$ (purple curve) with $\boldsymbol{\theta}^{(1)} = \left(\lambda, \kappa, \sigma\right)^\top = \left(0.104, 101.754, \sigma  \right)^\top$.  Results in (c)--(d) illustrate the outcome after ten iterations, with (c) showing the updated spline (red curve) superimposed on the data (blue dots) and (d) showing $\xi(t;\boldsymbol{\theta}^{(10)})$ (purple curve) with $\boldsymbol{\theta}^{(10)} = (\lambda, \kappa, \sigma)^\top = (0.108, 99.658, 3.841)^\top$.  Calculations are performed using $J=11$ and $K=1001$.\label{fig:F4}}
\end{figure}

The exact solution of Equation~\eqref{eq:LogisticODE} is
\begin{equation}
C(t)=\dfrac{\kappa C(0)}{C(0)+\left(\kappa-C(0)\right)\textrm{e}^{-\lambda t}}.
\end{equation}
Rather than use this closed form, we estimate parameters via generalized profiling, as in Section~\ref{sec: Newton}, using synthetic data with additive Gaussian noise\footnotemark[2]\footnotetext[2]{In Section~\ref{sec: CRN} we relax this assumption and consider alternative noise models.}. Figure~\ref{fig:F4}(a) uses $(\lambda,\kappa,C(0),\sigma)^\top=(0.1,100,5,5)^\top$ with observations at $t=0,10,20,\ldots,100$. We estimate $\boldsymbol{\theta}=(\lambda,\kappa,\sigma)^\top$ and calculate $C(0)$ afterwards. The data show near-exponential growth with associated time-scale $1/\lambda$ initially, and then fluctuate around an approximately constant level after $t\approx 60$.

The first step in generalized profiling is to over-fit the noisy data with a B-spline $f(t)$, as illustrated in Figure~\ref{fig:F4}(a). This B-spline is obtained following the same procedure outlined in Section~\ref{sec:Spline}. With the spline $f(t)$ we compute    
\begin{equation}
\xi(t; \boldsymbol{\theta}) = \dfrac{\textrm{d}f(t)}{\textrm{d}t} - \lambda \left[1  - \dfrac{f(t)}{\kappa}\right],
\end{equation}
and use numerical optimization to estimate $\boldsymbol{\theta} = (\lambda, \kappa, \sigma)^\top$ which maximizes $\ell_{\textrm{m}}$, given by Equation~\eqref{eq:modelerror}. Again, we evaluate $\ell_{\textrm{m}}$ using a uniform discretization of $t \in [0, 100]$ with $K=1001$ mesh points. This first optimization gives  $\boldsymbol{\theta}^{(1)} = (\lambda, \kappa, \sigma)^\top = (0.104, 101.754, \sigma)^\top$, noting again this first optimization step does not provide information about $\sigma$. As before, we use $\xi(t; \boldsymbol{\theta})$ to provide a quantitative measure of the degree to which the spline $f(t)$ fails to satisfy the governing ODE. In particular, the plot of $\xi(t; \boldsymbol{\theta}^{(1)})$ in Figure~\ref{fig:F4}(b) confirms that the spline in Figure~\ref{fig:F4}(a) does not satisfy Equation~\eqref{eq:LogisticODE} since $\xi(t; \boldsymbol{\theta}^{(1)})$ is relatively far from zero. Here, the poor match with is consistent with our intuitive expectations: we know in advance that the solution of Equation~\eqref{eq:LogisticODE} for $C(0)/\kappa \ll1$ is  monotonically increasing and bounded above by $\kappa$, so that $C(t) \to \kappa^{-}$ as $t \to \infty$. Here we see that the spline in Figure \ref{fig:F4}(a) violates both these properties.

We iteratively update the spline using the same algorithm outlined in Section~\ref{sec: Newton}.  After ten iterations we obtain the spline in Figure~\ref{fig:F4}(c).  Visually we see that the updated spline matches the data in a least--squares sense, and appears to capture the expected properties that $f(t)$ is monotonically increasing and bounded above by $\kappa$, so that $f(t) \to \kappa^{-}$ as $t$ becomes sufficiently large.  Our parameter estimates are $\boldsymbol{\theta}^{(10)} = (\lambda, \kappa, \sigma)^\top = (0.108, 99.658, 3.841)^\top$, and the corresponding plot of $\xi(t; \boldsymbol{\theta}^{(10)})$ in Figure~\ref{fig:F4}(d) confirms that the updated spline approximately enforces the ODE, and the updated spline gives $C(0) = f(0) = 3.936$.

\subsection{Modeling with systems of ODEs}\label{sec: CRN}

In this section we consider a further synthetic data example with two main learning objectives. The first objective is to work with a system of coupled ODEs, illustrated using a simple chemical reaction network describing the sequential decay of a parent chemical with concentration $C_1(t)\ge 0$ into a daughter chemical with concentration $C_2(t)\ge 0$. Despite its apparent simplicity, this chemical reaction network has applications in environmental pollution and wastewater treatment, such as modeling nitrification--denitrification processes, where ammonium is sequentially converted into nitrite and then nitrate, $\mathrm{NH}_4^{+}\rightarrow \mathrm{NO}_2^{-}\rightarrow \mathrm{NO}_3^{-}$. Focusing on the first two chemical species, linear mass action gives~\citep{Cho1971,Lunn1996}
\begin{subequations}\label{eq:CoupledODE}
\begin{align}
\frac{\mathrm{d}C_1(t)}{\mathrm{d}t} &= - r_1 C_1(t), \label{eq:CoupledODEa}\\
\frac{\mathrm{d}C_2(t)}{\mathrm{d}t} &= r_1 C_1(t) - r_2 C_2(t), \label{eq:CoupledODEb}
\end{align}
\end{subequations}
where $C_1(t)\ge 0$ is the concentration of ammonium $\mathrm{NH}_4^{+}$, $C_2(t)\ge 0$ is the concentration of nitrite $\mathrm{NO}_2^{-}$, $r_1>0$ is the decay rate of ammonium into nitrite, and $r_2>0$ is the decay rate of nitrite into nitrate. This model can be extended by adding a third ODE for nitrate, $C_3(t)\ge 0$, but the first two equations are sufficient to illustrate the key ideas. While we motivate this coupled model in terms of nitrification--denitrification, the same model is used to describe sequential radioactive decay~\citep{Bateman1910} and cell differentiation~\cite{Takahashi2021}, while nonlinear extensions of the model have been used to describe cell proliferation~\cite{Simpson2020Practical} with a carrying capacity density similar to the classic logistic growth model covered in Section~\ref{sec: Logistic}. 

\begin{figure}[htp]
  \centering
\includegraphics[width=1.00\textwidth]{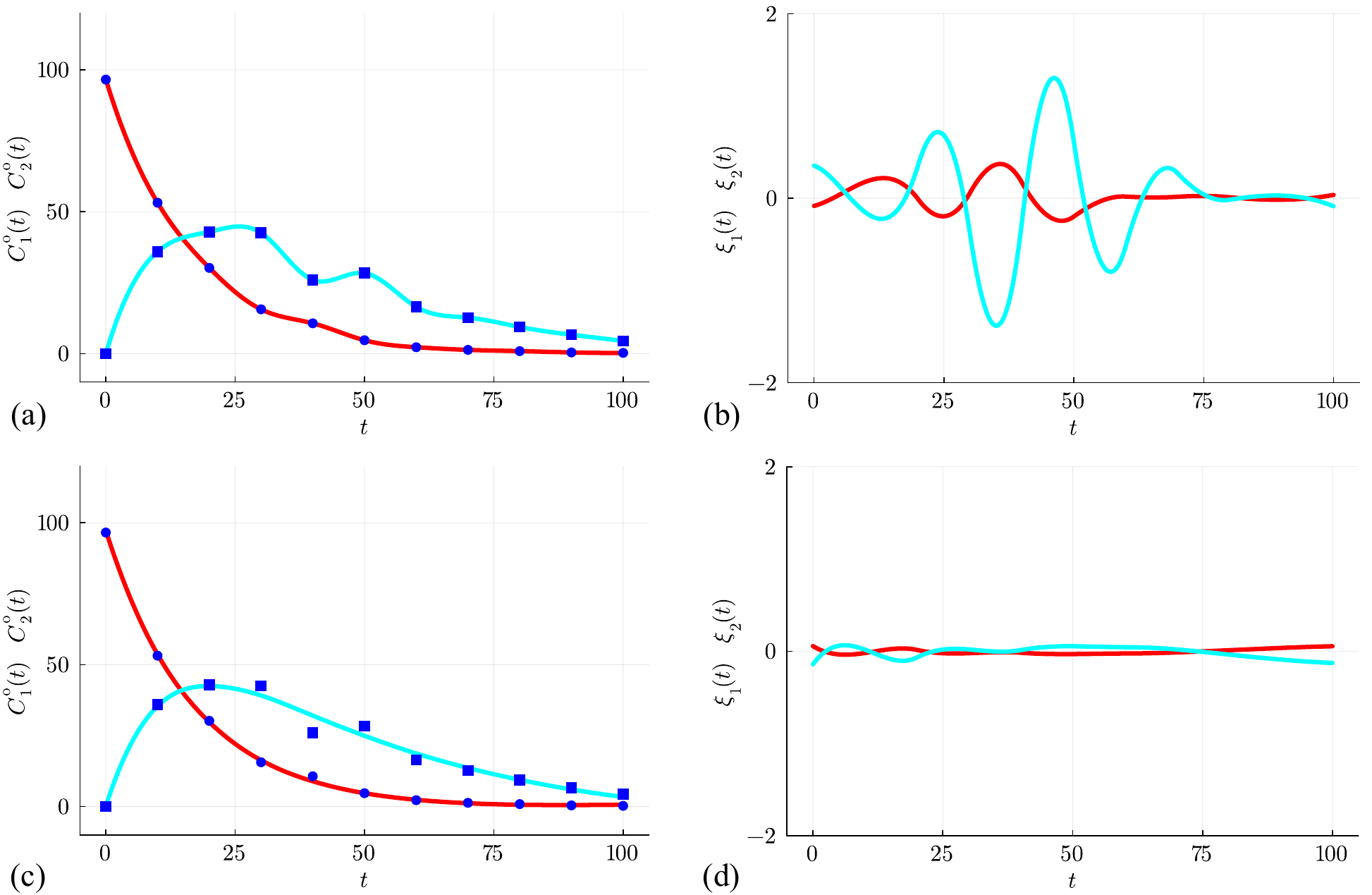}
  \caption{CRN case study with synthetic data. (a) Data for $C_1^{\textrm{o}}(t)$ (blue dots) and $C_2^{\textrm{o}}(t)$ (blue squares) are collected at $t=0, 10, 20,\ldots, 100$ and interpolated using over-fitted splines $f_1(t)$ (red curve) and $f_2(t)$ (cyan curve). (b) Plot of $\xi_1(t;\boldsymbol{\theta}^{(1)})$ (red curve) and $\xi_2(t;\boldsymbol{\theta}^{(1)})$ (cyan curve) with $\boldsymbol{\theta}^{(1)} = (r_1, r_2, \sigma )^\top = (0.061,0.041, \sigma)^\top$.  Results in (c)--(d) illustrate the outcome after ten iterations, with (c) showing the updated splines $f_1(t)$ (red curve) and $f_2(t)$ (cyan curve) superimposed on the same data. (d) Estimates of $\xi_1(t;\boldsymbol{\theta}^{(10)})$ (red curve) and $\xi_2(t;\boldsymbol{\theta}^{(10)})$ (cyan curve) with $\boldsymbol{\theta}^{(10)} = (r_1, r_2, \sigma )^\top = (0.059,0.040,0.238)^\top$. Calculations are performed using $J=11$ and $K=1001$. \label{fig:F5}}
\end{figure}

The second objective is motivated by noting the long-term solution of the system~\eqref{eq:CoupledODE} has $C_1(t)\to 0^+$ and $C_2(t)\to 0^+$ as $t\to\infty$ for all choices of $r_1$, $r_2$, $C_1(0)$, and $C_2(0)$. This property, which can be proved in the phase plane~\cite{murray2002}, strongly suggests that an additive Gaussian noise model with constant variance is inappropriate because it implies that measurements of $C_1(t)$ and $C_2(t)$ can be negative, especially at late time~\citep{Simpson2026,Murphy2024}. More broadly, the choice of the noise is an important consideration, and it depends on the modeling context. For example, in Section~\ref{sec: Newton} we consider a model of temperature where the lowest possible physical bound is absolute zero ($-273.15~^\textrm{o}\mathrm{C}$), which is irrelevant in many practical situations such as in Figure \ref{fig:F2} where we consider cooling from approximately $180~^\textrm{o}\mathrm{C}$ to $20~^\textrm{o}\mathrm{C}$. In contrast, when the dependent variable represents a count of objects~\citep{Simpson2024} or a density, such as in Equation~\eqref{eq:LogisticODE} and Equation~\eqref{eq:CoupledODE}, it is important to remember that counts and densities are always non-negative. One way to address this is to use a different noise model, such as a multiplicative noise model~\citep{Murphy2024}. An additional attraction of working with multiplicative noise is that the magnitude of the fluctuations are non-constant~\citep{Simpson2026,Murphy2024}.  For this example we generate the $j$th synthetic data point as follows, $C_1^{\rm{o}}(t_j)  = C_1(t_j) \eta$ and $C_2^{\rm{o}}(t_j) = C_2(t_j)\eta$  where $\eta~\sim~\textrm{Log-Normal}(0,\sigma^2)$~\citep{Murphy2024}.  In Figure~\ref{fig:F5}(a) we plot synthetic data corrupted with multiplicative log-normal noise for $(r_1, r_2, C_1(0), C_2(0),\sigma)^\top = (0.06,0.04,100,0,0.1)^\top$ noting that the fluctuations vanish as $t$ becomes sufficiently large and  $C_1(t) \to 0^+$ and $C_2(t) \to 0^+$. Data are collected for both chemical species  at $t=0, 10, 20,  \ldots, 100$, so that $J=11$. 

Taking a traditional approach to parameter estimation involves working with the solution of the system \eqref{eq:CoupledODE}.  The solution can be obtained using several techniques, such as Laplace transforms~\cite{Bateman1910}, and can be written as
\begin{subequations}\label{eq:CoupledODESolution}
\begin{align}
C_1(t) &= C_1(0)\mathrm{e}^{-r_1 t}, \label{eq:CoupledODESolutiona}\\
C_2(t) &=
\left\{
\begin{aligned}
& C_2(0)\mathrm{e}^{-r_2 t}
 - r_1 C_1(0)\left[\frac{\mathrm{e}^{-r_1 t}-\mathrm{e}^{-r_2 t}}{r_1-r_2}\right],
&& r_1 \neq r_2, \label{eq:CoupledODESolutionb}\\
& \big(C_2(0)+ r_1 C_1(0)\, t\big)\mathrm{e}^{-r_1 t},
&& r_1 = r_2.  
\end{aligned}
\right.
\end{align}
\end{subequations}
The computational implementation of this solution requires care since the expression for $C_2(t)$ takes a different algebraic form depending on the values of $r_1$ and $r_2$.  Standard approaches for parameter estimation (e.g.~MLE, MCMC) require the use of different expressions for $C_2(t)$ which can become challenging because the expression for $C_2(t)$ and $r_1 \ne r_2$ is ill-conditioned in the region of parameter space for which $|r_1 - r_2|$ is sufficiently small.  Evaluating the solution in this region of the parameter space can introduce significant floating-point round-off and cancellation errors.  Beyond dealing with this algebraic complication, standard parameter estimation approaches (e.g.~MLE, MCMC) target $\boldsymbol{\theta}_{\textrm{full}} = (r_1, r_2, C_1(0),C_2(0), \sigma)^\top$, since estimates of $C_1(0)$ and $C_2(0)$ are required to evaluate the exact solution, Equation~\eqref{eq:CoupledODESolution}.   In contrast, generalized profiling completely avoids both complications, and we simply estimate $\boldsymbol{\theta} = (r_1, r_2, \sigma)^\top$ without the need for considering special cases\footnotemark[1]\footnotetext[1]{The algebraic complication can also be avoided by working with a numerical solution of the system~\eqref{eq:CoupledODE}. This approach introduces additional computational overhead in solving the system of ODEs numerically, and still requires the estimation of $C_1(0)$ and $C_2(0)$.}.

We first use two B-splines, $f_1(t)$ and $f_2(t)$, to over-fit the data in Figure~\ref{fig:F5}(a). With $f_1(t)$ and $f_2(t)$ we compute
\begin{subequations}
\begin{align}\label{eq:Coupledspline}
\xi_1(t; \boldsymbol{\theta}) &= \dfrac{\textrm{d}f_1(t)}{\textrm{d}t} +r_1 f_1(t), \\
\xi_2(t; \boldsymbol{\theta}) &=\dfrac{\textrm{d}f_2(t)}{\textrm{d}t}- r_1 f_1(t) + r_2 f_2(t).
\end{align}
\end{subequations}
To accommodate working with a system of ODEs, we extend our definition of $\ell_{\textrm{m}}$ to give
\begin{equation}
\ell_{\textrm{m}}\left(\boldsymbol{\theta}; C_1^{\textrm{o}}(t), C_2^{\textrm{o}}(t) \right) = -\dfrac{1}{K}\sum_{s=1}^{S}\sum_{k=1}^{K}\left[\xi_{s}(t_k;\boldsymbol{\theta})\right]^2, \label{eq:modelerror2}
\end{equation}
where $S=2$ in this case since measurements are made for two species. Using numerical optimization we compute $\boldsymbol{\theta} = (r_1, r_2, \sigma)^\top$ that maximizes $\ell_{\textrm{m}}$, giving $\boldsymbol{\theta}^{(1)} = (r_1, r_2, \sigma )^\top = (0.061,0.041, \sigma)^\top$, noting again that $\ell_{\textrm{m}}$ is independent of $\sigma$ so that we do not obtain any estimate of $\sigma$ in this first step.  With these parameter estimates we evaluate and plot $\xi_1(t;\boldsymbol{\theta}^{(1)})$ and $\xi_2(t;\boldsymbol{\theta}^{(1)})$ in Figure~\ref{fig:F5}(b), confirming that the splines in Figure~\ref{fig:F5}(a) do not accurately enforce the ODEs.  We continue with the same approach as in Sections \ref{sec: Newton}--\ref{sec: Logistic} with the exception that we use an appropriately modified log-likelihood,
\begin{equation}
\ell_{\textrm{d}}\left(\boldsymbol{\theta}; C_1^{\textrm{o}}(t),C_2^{\textrm{o}}(t) \right) = \sum_{s=1}^{S}\sum_{j=1}^{J} \log \left[\phi\left(C_{s}^{\textrm{o}}(t_j); \log \left(f_{s}(t_j)\right), \sigma^2  \right) \right], 
\end{equation}
where $\phi(x; \mu, \sigma^2)$ is the density of the Log-Normal$(\mu,\sigma^2)$ distribution.  With these ingredients we iteratively update the spline coefficients, parameters and weights as before. After 10 iterations we obtain the splines in Figure~\ref{fig:F5}(c) and the associated parameter estimates $\boldsymbol{\theta}^{(10)} = (r_1, r_2, \sigma )^\top = (0.059,0.040,0.238)^\top$. Plots of $\xi_1(t;\boldsymbol{\theta}^{(10)})$ and $\xi_2(t;\boldsymbol{\theta}^{(10)})$  in Figure~\ref{fig:F5}(d) confirm the updated splines approximately enforce the coupled ODEs relative to the initial splines in Figure~\ref{fig:F5}(a). Using the splines in Figure~\ref{fig:F5}(c) we obtain $C_1(0) = f_1(0) = 97.281$ and $C_2(0) = f_2(0) = 0.170$, which compare reasonably well with the true values of $C_1(0)=100$ and $C_2(0) = 0$. Overall, the key practical difference in working with the ODE system is that Equation \eqref{eq:LeastSquares} is expanded to incorporate additional data matching and model matching rows for both $C_1(t)$ and $C_2(t)$.  The remainder of the algorithm proceeds unchanged.

Before concluding this section it is worth briefly commenting that the relative advantages of generalized profiling become even more compelling for larger systems of ODEs. For example, consider the implications of extending the system~\eqref{eq:CoupledODE} to include an additional chemical species. In the context of an extended model of nitrification-denitrification~\citep{Cho1971,Lunn1996} we could write
\begin{subequations}\label{eq:CoupledODE2}
\begin{align}
\frac{\mathrm{d}C_1(t)}{\mathrm{d}t} &= - r_1 C_1(t), \label{eq:CoupledODEa2}\\
\frac{\mathrm{d}C_2(t)}{\mathrm{d}t} &= r_1 C_1(t) - r_2 C_2(t), \label{eq:CoupledODEb2}\\
\frac{\mathrm{d}C_3(t)}{\mathrm{d}t} &= r_2 C_2(t) - r_3 C_3(t), \label{eq:CoupledODEb3}
\end{align}
\end{subequations}
where $C_3(t)\ge0$ is the concentration of nitrate, and $r_3 >0$ is the rate of nitrate decay~\citep{Cho1971,Lunn1996}. In this modestly extended framework, traditional parameter estimation approaches require the solution of the system~\eqref{eq:CoupledODE2}. As before, evaluating an exact solution of the model requires us to consider special cases ($r_1=r_2$, $r_2=r_3$, $r_1=r_3$), whereas obtaining numerical solutions of the system introduces an additional computational overhead relative to evaluating the exact solution.  Regardless of how the ODE system is solved, working within a traditional parameter estimation framework (e.g.~MLE, MCMC) also requires the simultaneous estimation of $\boldsymbol{\theta}_{\textrm{full}} = (r_1, r_2, r_3, C_1(0),C_2(0), C_3(0), \sigma)^\top$.  Generalized profiling avoids both complications since there are no special cases to consider and instead of targeting seven unknowns we would target four, $\boldsymbol{\theta} = (r_1, r_2, r_3, \sigma)^\top$.

\subsection{Practical application: Coral reef recovery data}\label{sec:Coral}

\begin{figure}[htp]
  \centering
\includegraphics[width=1.0\textwidth]{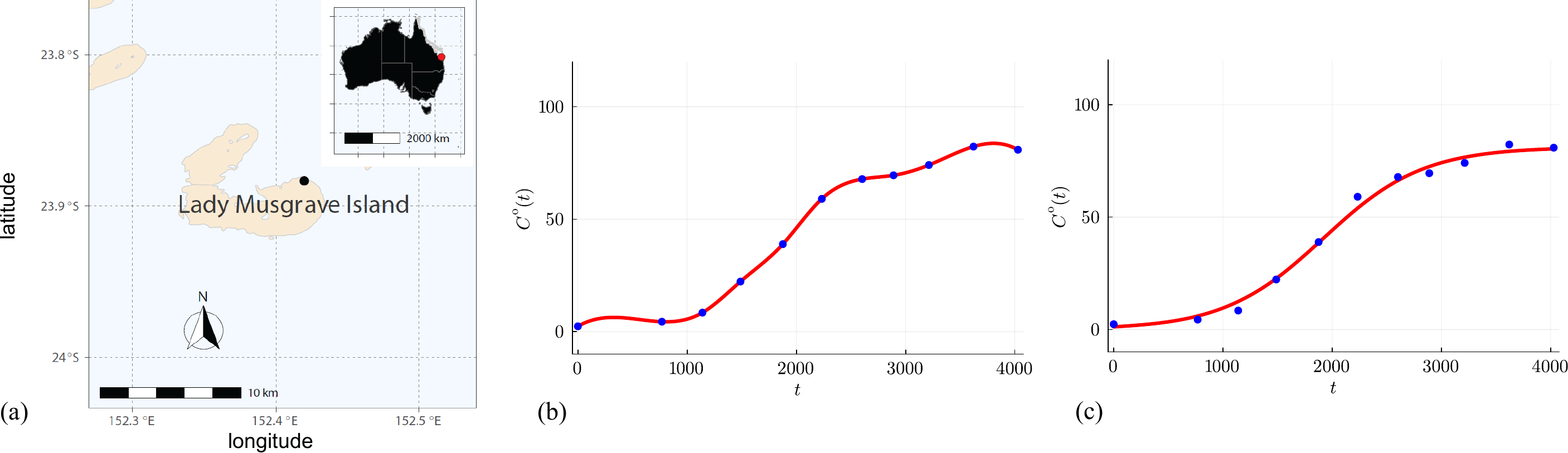}
  \caption{Logistic growth case study with field data. (a) Location of Lady Musgrave Island (black dot) off the East coast of Australia at the Southern end of the Great Barrier Reef (inset, red dot). (b) Field data (blue dots) showing the percentage area covered by hard corals as a function of time after some external disturbance. The data is superimposed with an over-fitted spline (red curve).  (c)  Field data (blue dots) superimposed with the updated spline (red curve) after 10 iterations giving $\boldsymbol{\theta} = (\lambda, \kappa, \sigma)^\top = (2.20 \times 10^{-3}, 81.19, 2.58)^\top$. \label{fig:F6}}
\end{figure}

We conclude this tutorial by applying the ideas to a practical data set.  Data in Figure~\ref{fig:F6}(b) summarize measurements of the percentage of hard coral cover on a coral reef near Lady Musgrave Island on the Australian Great Barrier Reef whose location is shown in Figure~\ref{fig:F6}(a)~\citep{Simpson2022,Warne2022,Simpson2023}.  This data  describes the 11-year-long recovery in hard coral cover after some major disturbance (e.g.~tropical cyclone or heat related bleaching) causes a sudden reduction in coral cover from which we are interested to follow and model the recovery trajectory.  The data shows a classic sigmoid growth curve.   

The main motivation for analysing these data with a mechanistic model is to quantify how quickly coral communities recover from disturbances, and whether recovery can keep pace with increasingly frequent climate-driven events~\citep{Simpson2022}. Model-based analysis yields biologically meaningful estimates (e.g., growth rates, carrying capacities, recovery time scales) and enables quantitative comparisons across different data sets, such as comparing recovery rates and recovery trajectories from different reefs and disturbance histories.   

We interpret the data in Figure \ref{fig:F6}(b) using the logistic growth model, Equation \eqref{eq:LogisticODE}, by making the standard assumption that the noisy data are described by the solution of the logistic growth model corrupted with additive Gaussian noise with constant variance.  Accordingly, we will estimate $\boldsymbol{\theta} = (\lambda, \kappa, \sigma)^\top$ using generalized profiling, noting that the estimate of $\lambda$ provides a simple measure of the recovery time-scale $1/\lambda$, and $\kappa$ provides an estimate of the long-term total coral biomass.  For the synthetic data cases in Section \ref{sec: Newton}--\ref{sec: CRN} we have not mentioned any units or dimensions of the dependent variables, independent variables, or parameters.  Here we work with data that characterizes the coral cover in terms of percentage area cover, and the measurements of time are made in units of days.  This means that the units of $\lambda$ are 1/day, while the units of $\kappa$ and $\sigma$ are percentage area cover.

Unlike the synthetic data sets in Sections~\ref{sec: Newton}--\ref{sec: CRN}, the coral reef recovery data are collected under real-world constraints over an 11-year-long time period.  A consequence of these constraints is that the data are not uniformly spaced.  Fortunately, interpolating this data with B-splines remains practically unchanged relative to working with uniformly spaced data. In this case we have data collected over 4028 days, with $J=11$ measurements at $t=[0, 769, 1140, 1488, 1876, 2233, 2602, 2889, 3213, 3621, 4028]$.  When we apply generalized profiling we enforce the governing ODE on a relatively fine mesh, $t=0, 1, 2, 3, \ldots, 4027,4028$, which means that  we use $K=4029$ points to enforce the ODE.

Results in Figure~\ref{fig:F6}(b) show the data superimposed on the first over-fitted spline.  The steps in the generalized profiling procedure are the same as those followed in Section~\ref{sec: Logistic}, and can be followed in the Jupyter notebooks.  The over-fitted spline in Figure~\ref{fig:F6}(b) is inconsistent with the qualitative features of the logistic growth model in exactly the same way as discussed in Section~\ref{sec: Logistic} for the synthetic data exercise.  The updated spline after 10 iterations in Figure~\ref{fig:F6}(c) has properties that are consistent with the logistic model, leading to estimates $\boldsymbol{\theta} = (\lambda, \kappa, \sigma)^\top = (2.20 \times 10^{-3}, 81.19, 2.58)^\top$.  These estimates compare very well with previous estimates obtained using the same data set, the same mathematical model, and the same additive Gaussian noise model, except that these previous parameter estimates are obtained using MLE~\citep{Simpson2022}.  As with the synthetic data examples in Sections~\ref{sec: Newton}--\ref{sec: CRN}, we can use our spline approximation to estimate $C(0)$, giving $f(0) = C(0) = 1.18$.  To keep our presentation succinct we have omitted plots of $\xi(t;\boldsymbol{\theta}^{(1)})$ and $\xi(t;\boldsymbol{\theta}^{(10)})$ in Figure~\ref{fig:F6} because these plots are very similar to those shown in Figures \ref{fig:F2}--\ref{fig:F5}.  We encourage readers to explore the Jupyter notebooks on GitHub where plots of $\xi(t;\boldsymbol{\theta}^{(n)})$ for $n=1,2,3,\ldots,10$ are provided.

\section{General remarks and future opportunities} \label{sec:conclusion}

This tutorial provides a set of self-contained computational exercises to develop skills in parameter estimation for ODE models using generalized profiling (parameter cascading). Rather than repeatedly solving the ODE (as in MLE or ABC), the method fits splines to noisy data and then iteratively updates the splines to balance data fit and model fit. Open-source Jupyter notebooks on GitHub enable the replication of all calculations, as well as providing a platform to extend the algorithm to other ODE models.

There are many ways to extend the tools and investigations presented in this tutorial. There are a range of minimal but instructive extensions that simply involve varying implementation details as follows. Most exercises here use data at $J=11$ time points and enforce the model at $K=1001$ time points to provide a balance between accuracy and computational cost.  Exploring the impact of varying $K$ is straightforward and insightful; smaller $K$ gives inaccurate parameter estimates while larger $K$ increases computational overhead. Another minimal extension is to repeat the calculations using natural boundary conditions for the splines. Both of these extensions are straightforward to implement in the open access notebooks provided, and we leave these ideas for future exercises to support further skills development.  

A more substantial extension is to apply generalized profiling to partial differential equation (PDE) models as this requires interpolation across more than one independent variables (e.g., using tensor-product splines)~\citep{Botella2003}.  Working with PDE models is of high interest since the computational savings in avoiding the need to solve the underlying mathematical model will be more significant for PDE models relative to ODE-based mathematical models.   Another important direction is to consider parameter identifiability, which is not addressed here because we chose to work with a range of problems that are known to be practically identifiable~\citep{Simpson2026}. In likelihood-based settings, practical identifiability is often assessed using some measure the curvature of the log-likelihood at the MLE, and Wilks' theorem is used to define asymptotic confidence regions~\citep{Casella2002,Wilks1938}.  In sampling-based settings, such as likelihood-based MCMC, non-identifiability is usually characterized by poor chain mixing and diffuse posteriors~\citep{Hines2014,Siekmann2012,Simpson2020Practical}. For penalized likelihood, standard asymptotic thresholds may not apply but progress can be made computationally using bootstrap methods~\citep{Frohlich2014}. An interesting and relevant extension is to apply generalized profiling to structurally and/or practically non-identifiable problems and explore how non-identifiability affects generalized profiling~\citep{Frohlich2014,SimpsonMaclaren2024PoorlyIdentified}.

As we mentioned in the Introduction, generalized profiling has similar aims to physics-informed neural networks (PINNs), but uses a different approximation and computational strategy. In a PINN, the unknown solution of a particular differential equation is approximated by a neural network. The network is trained by minimizing a loss function that measures both disagreement with the observed data and failure to satisfy the differential equation~\cite{Raissi2019}. When the differential equation contains unknown parameters, these parameters can be estimated together with the neural-network weights. Biologically informed neural networks (BINNs) are often interpreted as providing an extension of this approach by using neural networks to represent unknown functions within the differential equation, such as a density-dependent growth-rate function~\cite{Lagergren2020BINNs}.  Generalized profiling addresses a closely related problem, but replaces the neural network with a B-spline approximation.  For fixed model parameters, the spline coefficients can be updated using least-squares calculations, after which the model parameters can be efficiently updated in a separate optimization step.  Repeating these two steps iteratively provides an efficient way to balance agreement with the data and agreement with the differential equation. For the low-dimensional ODE models considered in this tutorial article, the computational approach is relatively inexpensive, straightforward to implement, and requires fewer computational choices and specification of hyperparameters than a neural network, for which the user must select an architecture, activation functions, optimization algorithm, learning rates, and other training settings~\cite{Raissi2019}.  PINNs and BINNs are nevertheless more flexible in settings involving high-dimensional models or unknown functional relationships.  Overall, generalized profiling, PINNs, and BINNs can be interpreted as related approaches that aim to combine data with differential equations, with each method offering different advantages depending on the problem under consideration.

\noindent
\textbf{Acknowledgments.} M.J.S. is supported by the Australian Research Council (CE230100001, DP230100025).  R.E.B. is supported by a grant from the Simons Foundation (MP-SIP-00001828). For the purpose of open access, the author has applied a CC BY public copyright licence to any author accepted manuscript arising from this submission.

\noindent
\textbf{Data Accessibility.} Open source Julia software is available on \href{https://github.com/ProfMJSimpson/PenalizedLikelihood}{GitHub}.  Data used in the case study is available in the Julia software provided.

\newpage
\printbibliography
%\bibliography{references}

\end{document}